\documentclass[aps,prl,reprint,twocolumn,superscriptaddress]{revtex4-1}
\usepackage{amsmath}
\usepackage{amsfonts}
\usepackage{amssymb}
\usepackage{amsthm}
\usepackage{graphicx}
\usepackage{CJK}
\usepackage{times}
\usepackage{mathptmx}
\usepackage[colorlinks, linkcolor=blue, urlcolor=blue, anchorcolor=blue, citecolor=blue]{hyperref}
\usepackage{mathtools}
\usepackage{lipsum}
\usepackage{array}
\usepackage[table]{xcolor}
\definecolor{mygray}{gray}{.9}

\begin{document}


\title{Symmetry breaking induced magnon-magnon coupling in synthetic antiferromagnets}



\author{Jie Lu}
\affiliation{College of Physics and Hebei Advanced Thin Films Laboratory, Hebei Normal University, Shijiazhuang 050024, People's Republic of China}
\author{Mei Li}
\email{limeijim@163.com}
\affiliation{Physics Department, Shijiazhuang University, Shijiazhuang, Hebei 050035, People's Republic of China}
\author{Wei He}
\email{hewei@iphy.ac.cn}
\affiliation{State Key Laboratory of Magnetism and Beijing National Laboratory for Condensed Matter Physics, Institute of Physics, Chinese Academy of Sciences, Beijing 100190, People's Republic of China}


\date{\today}

\begin{abstract}
We propose a general theory of microwave absorption spectroscopy for symmetry-breaking 
synthetic antiferromagnets (SAFs).
Generally, inhomogeneity or different thickness of the two ferromagnetic sublayers of a SAF results in the 
intrinsic symmetry breaking, while out-of-plane components of dc magnetic fields lead to the extrinsic one.
The broken symmetry of SAFs excludes the original symmetry-protected
crossing between pure in-phase and out-of-phase resonance modes with opposite parity.
Alternatively, new frequency branches become hybridization of original bare modes in terms of
symmetry-breaking-induced magnon-magnon coupling, which results in an indirect gap 
in ferromagnetic resonance frequencies.
Also, the dependence of gap width on the degree of symmetry breaking for several typical cases 
are presented and compared with existing experiments.
Our theory provides a simple but physical understanding on the rich structure of ferromagnetic resonance
spectra for asymmetric SAFs.

\end{abstract}


\maketitle


Synthetic antiferromagnets (SAFs) are magnetic multilayers with two ferromagnetic (FM) sublayers 
coupled antiferromagnetically through a nonmagnetic metallic spacer\cite{Duine_nphys_2018}.
They have attracted tremendous interest in the past decades due to
their potential for developing the ``SAF spintronics"  and wide range of applications in magnetic nanodevices
\cite{Seki_APL_2009,Gonzalez_PRB_2013,Timopheev_PRB_2014,Tanaka_APExpress_2014,Yang_APL_2016,Li_AFM_2016,CWang_APL_2018,WWang_APL_2018,XFHan_APL_2018,Kamimaki_APL_2019,Chen_APL_2019,Sorokin_PRB_2020,Kamimaki_PRAppl_2020,Martin_JAP_2007,Belmeguenai_JPCM_2008,Liu_PRB_2014,Shiota_PRL_2020,Shiota_SciAdv_2020,Sud_PRB_2020,Waring_PRApplied_2020}.
Compared with the strong exchange coupling in genuine antiferromagnetic (AFM) materials with terahertz intrinsic 
frequencies\cite{Kampfrath_nphoton_2011,Baierl_PRL_2016}, the relatively weak interlayer coupling in SAFs
mainly comes from the Ruderman-Kittel-Kasuya-Yosida (RKKY) interaction\cite{Kittel_PhyRev_1954,Yafet_JAP_1987,Bruno_PRL_1991}
and thus realizes gigahertz FM resonance (FMR) frequencies that mature microwave electronics can match.
Similar behaviors have also been observed in layered crystals\cite{MacNeill_PRL_2019,XiaoxiaoZhang_nmat_2020}
and compensated ferrimagnets\cite{Liensberger_PRL_2019} with AFM inter-layer
and inter-sublattice couplings, respectively.
More interestingly, for symmetrical NiFe/Ru/NiFe\cite{Martin_JAP_2007,Belmeguenai_JPCM_2008} and
FeCoB/Ru/FeCoB SAFs\cite{Shiota_PRL_2020,Shiota_SciAdv_2020,Sud_PRB_2020,Waring_PRApplied_2020},
or layered crystal $\mathrm{CrCl_3}$\cite{MacNeill_PRL_2019}, 
symmetry-protected mode crossings between in-phase and out-of-phase branches of 
FMR spectra have been observed under in-plane external dc magnetic fields.
This indicates the absence of coupling between magnons with opposite parity, 
which is, on the contrary, very common in yttrium iron garnet/ferromagnet bilayers\cite{JChen_PRL_2018,Klingler_PRL_2018,YLi_PRL_2020}.

In fact, this mode crossing can be eliminated in several ways. 
For symmetrical SAFs or layered crystals, extrinsically exerting an out-of-plane dc field will lift
the system's rotation-symmetry axis away from the SAF plane, thus
breaks the rotation symmetry of the hard axis (normal of SAF plane) from the magnetostatic interaction.
This introduces a magnon-magnon coupling between the original uncoupled modes with opposite parity,
hence hybridizes the two modes and generate an anticrossing gap\cite{Sud_PRB_2020,MacNeill_PRL_2019}.
Very recently, strong magnon-magnon coupling under in-plane
dc fields is also proposed by the dynamic dipolar interaction (nonuniform precession) 
in symmetrical FeCoB/Ru/FeCoB SAFs\cite{Shiota_PRL_2020,Shiota_SciAdv_2020}, 
which provides an alternative way of extrinsic symmetry breaking (SB).
The other strategy is to break the intrinsic symmetry between the two FM sublayers in SAFs.
In most existing experiments the two sublayers are prepared from the same FM materials but 
with different thickness. 
A frequency gap can be observed even under in-plane dc fields\cite{Belmeguenai_JPCM_2008,Liu_PRB_2014}.
In addition, intrinsic asymmetry should also appear when the two sublayers are made from different FM materials.
However, to our knowledge the corresponding FMR measurements are not yet in the press,
which mainly comes from the difficulty in sample preparation.

A lot of theoretical work has been performed to understand the rich structure of FMR spectra in SAFs\cite{Gonzalez_PRB_2013,Belmeguenai_JPCM_2008,Liu_PRB_2014,MacNeill_PRL_2019,Liensberger_PRL_2019,Shiota_PRL_2020,Sud_PRB_2020,Waring_PRApplied_2020,Sorokin_PRB_2020,Kamimaki_PRAppl_2020}. 
Representatively, in 2014 a discrete-lattice approach 
has been raised for asymmetric NiFe/Ru/NiFe SAFs\cite{Liu_PRB_2014}, 
where the RKKY interaction, biquadratic exchange couplings and 
the uniaxial anisotropy at the NiFe/Ru interfaces are all considered,
however a clear and simple physical picture is still lacking.
In 2019, based on ``macrospin" assumption MacNeill \textit{et. al.} proposed a systematic 
analysis for the gap induction by extrinsic SB from out-of-plane dc fields
in which only the bilinear RKKY interaction is included\cite{MacNeill_PRL_2019}.
However, they did not consider the intrinsic SB since the sublayers of their 
system are always symmetrical.
In this Letter, we demonstrate that intrinsic and extrinsic SB in SAFs can induce
magnon-magnon couplings via different mechanisms,
thus further lead to indirect gaps in their FMR spectra.
In brief, a general theory including both intrinsic and extrinsic SB (from tilting dc fields) 
will be presented first and then followed by several examples that are easy to compare with experiments.

\begin{figure} [htbp]
	\centering
	\includegraphics[width=0.45\textwidth]{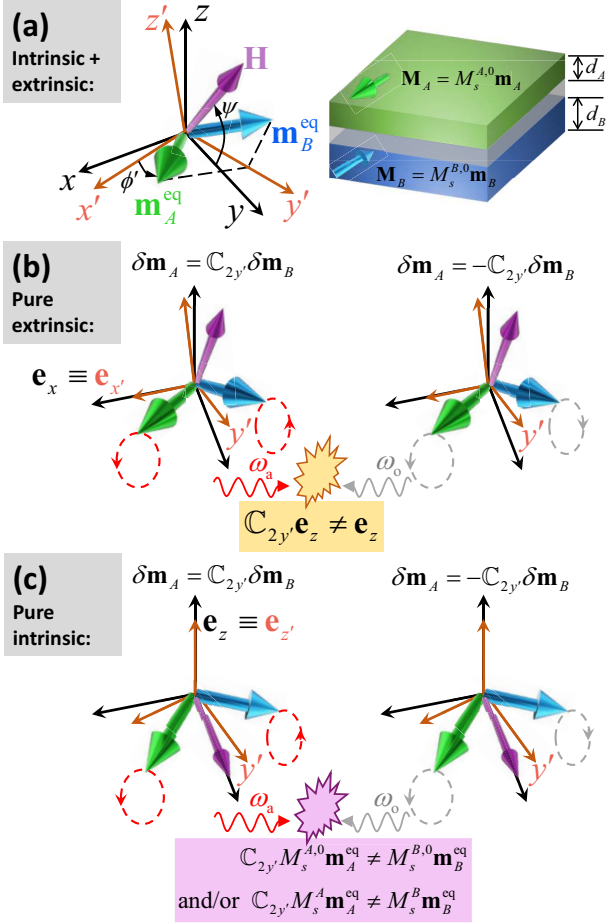}
	\caption{(a) Sketch of a typical SAF with saturation magnetization $M_s^{A(B),0}$
		and thickness $d_{A(B)}$ in its two FM sublayers. 
		The macrospins ($\mathbf{m}_A$ and $\mathbf{m}_B$)
		of the two sublayers are coupled antiferromagnetically and tilted at equilibrium
		 ($\mathbf{m}_A^{\mathrm{eq}}$ and $\mathbf{m}_B^{\mathrm{eq}}$)
		under both extrinsic and intrinsic SB. 
		Generally, the new $x'y'z'$ coordinate system (see definition
		in the main text) is totally different with the initial $xyz$ one. 
		The deviation of $z'$ ($x',y'$) from the $z$ ($x,y$) axis comes from the extrinsic (intrinsic) SB.
		(b) Sketch of magnon-magnon coupling between the in-phase
		($\delta\mathbf{m}_A=\mathcal{C}_{2y'}\delta\mathbf{m}_B$) 
		and out-of-phase ($\delta\mathbf{m}_A=-\mathcal{C}_{2y'}\delta\mathbf{m}_B$) 
		modes with opposite	parity under pure extrinsic SB with $\mathbf{e}_x\equiv\mathbf{e}_{x'}$.
		This coupling comes from the breaking rotation symmetry of magnetostatics under $\mathcal{C}_{2y'}$
		($\mathcal{C}_{2y'}\mathbf{e}_z\ne \mathbf{e}_z$).
		(c) Sketch of magnon-magnon coupling between in-phase and out-of-phase 
		modes under pure intrinsic SB with $\mathbf{e}_z\equiv\mathbf{e}_{z'}$.
		This coupling comes from the breaking rotation symmetry of magnetostatics and/or magnetization
		under $\mathcal{C}_{2y'}$ ($\mathcal{C}_{2y'}M_s^{A,0}\mathbf{m}_A^{\mathrm{eq}}\ne M_s^{B,0}\mathbf{m}_B^{\mathrm{eq}}$ and/or $\mathcal{C}_{2y'}M_s^{A}\mathbf{m}_A^{\mathrm{eq}}\ne M_s^{B}\mathbf{m}_B^{\mathrm{eq}}$).
	}\label{fig1}
\end{figure}

The SAF system under consideration is sketched in Fig. 1 (a). 
The saturation magnetization and thickness of the upper (lower) FM sublayer 
are $M_s^{A,0}$ and $d_A$ ($M_s^{B,0}$ and $d_B$), respectively.
The crystalline anisotropy in both sublayers are neglected for simplicity. 
In typical SAFs, the FM intralayer nearest-neighbor exchange
is much stronger than the interlayer AFM coupling.
Therefore the ``macrospin" approximation generally provides enough information about spin-wave behaviors.
Now the magnetization within each sublayer
can be viewed as uniform and denoted as $\mathbf{m}_A$ and $\mathbf{m}_B$, respectively.
The interlayer exchange energy is then modeled as $\mu_0 \lambda_{E} S (d_A+d_B) M_s^{A,0} M_s^{B,0} \mathbf{m}_A\cdot \mathbf{m}_B/2$, where $\lambda_{E}>0$ is a dimensionless coefficient and 
$S$ is the SAF cross-section area.
In literatures, this term is also expressed as the ``interlayer exchange energy per unit area"
$J_{\mathrm{IEC}}$\cite{Tanaka_APExpress_2014,Yang_APL_2016,WWang_APL_2018,Kamimaki_APL_2019,Kamimaki_PRAppl_2020,Shiota_PRL_2020,Shiota_SciAdv_2020,Waring_PRApplied_2020}. 
Hence $\lambda_{E}=2 J_{\mathrm{IEC}}/[\mu_0 M_s^{A,0}M_s^{B,0}(d_A+d_B)]$. 
By neglecting the Gilbert damping, a coupled Landau-Lifshitz-Gilbert (LLG) equation reads
\begin{equation}\label{Coupled_LLG_vectorial}
\begin{split}
\dot{\mathbf{m}}_A=&-\gamma\mathbf{m}_A\times\left[\mathbf{H}-\lambda_{E}M_s^B\mathbf{m}_B-M_s^{A,0}\left(\mathbf{m}_A\cdot\mathbf{n}\right)\mathbf{n}\right]+\mathbf{\tau}_A,  \\
\dot{\mathbf{m}}_B=&-\gamma\mathbf{m}_B\times\left[\mathbf{H}-\lambda_{E}M_s^A\mathbf{m}_A-M_s^{B,0}\left(\mathbf{m}_B\cdot\mathbf{n}\right)\mathbf{n}\right]+\mathbf{\tau}_B,
\end{split}
\end{equation}
where an overdot means $d/d t$, $\mathbf{n}$ is the sublayer normal and
\begin{equation}\label{MsA_MsB_definitions}
M_s^A=\frac{d_A+d_B}{2d_B}M_s^{A,0},\quad M_s^B=\frac{d_A+d_B}{2d_A}M_s^{B,0}
\end{equation}
are the ``thickness-modified saturation magnetization" of the respective sublayers.
$\gamma=\mu_0\gamma_e$ with $\mu_0$ and $\gamma_e$ being the vacuum permeability and 
electron gyromagnetic ratio, respectively.
At last, $\mathbf{\tau}_{A(B)}$ is the torque on $\mathbf{m}_{A(B)}$ which arises from the rf
excitation field of coplanar waveguides.

In the following we present a general theory describing the 
combining effects of intrinsic and extrinsic SB on FMR spectroscopy.
As preparation, we calculate the equilibrium magnetization orientations under a general dc magnetic field $\mathbf{H}$ with strength $H$ and tilting angle $\psi$ ($0\le \psi< \pi/2$).
First we construct the Cartesian coordinate system $(\mathbf{e}_x,\mathbf{e}_y,\mathbf{e}_z)$:
$\mathbf{e}_z\equiv\mathbf{n}$ thus the out-of-plane field component is
$H\sin\psi$, $\mathbf{e}_y$ is parallel to the in-plane field component $\mathbf{H}-H\sin\psi\mathbf{e}_z$,
and $\mathbf{e}_x=\mathbf{e}_y\times\mathbf{e}_z$.
Due to the AFM interlayer coupling, without $\mathbf{H}$ the unit magnetization vectors 
in the two sublayers orients oppositely in a certain in-plane direction in the absence of 
crystalline anisotropy.
When $\mathbf{H}$ is applied, in principle they are pulled out of $xy$ plane and their final equilibrium
states are denoted as $\mathbf{m}_A^{\mathrm{eq}}$ and $\mathbf{m}_B^{\mathrm{eq}}$, respectively.
We then set $\theta_{A(B)}$ as their respective polar angles, and 
$\phi_A$ ($\phi_B$) being the angle that the in-plane component of $\mathbf{m}_A^{\mathrm{eq}}$
($\mathbf{m}_B^{\mathrm{eq}}$)
rotates anticlockwise (clockwise) with respect to $\mathbf{e}_x$ ($-\mathbf{e}_x$).
These four angles can be explicitly solved from the static LLG equation (see Supplement Material).

Next we define a new $x'y'z'$ coordinate system based on 
$\mathbf{m}_A^{\mathrm{eq}}$ and $\mathbf{m}_B^{\mathrm{eq}}$:
$\mathbf{e}_{x'}\parallel \mathbf{m}_A^{\mathrm{eq}}-\mathbf{m}_B^{\mathrm{eq}}$, 
$\mathbf{e}_{y'}\parallel \mathbf{m}_A^{\mathrm{eq}}+\mathbf{m}_B^{\mathrm{eq}}$,
and $\mathbf{e}_{z'}\parallel \mathbf{m}_A^{\mathrm{eq}}\times\mathbf{m}_B^{\mathrm{eq}}$.
Note that to ensure the noncollinearity of $\mathbf{m}_A^{\mathrm{eq}}$ and $\mathbf{m}_B^{\mathrm{eq}}$,
the dc field strength $H$ should be limited within the range $H_{\mathrm{AFM}}<|H|<H_{\mathrm{FM}}$ 
(see Supplement Material for their general definitions).
For $|H|\le H_{\mathrm{AFM}}$ ($\ge H_{\mathrm{FM}}$), $\mathbf{m}_A^{\mathrm{eq}}$ and $\mathbf{m}_B^{\mathrm{eq}}$
point in the opposite (same) direction thus the SAF falls into AFM (FM) state.
Then we denote $\mathcal{C}_{2y'}$ as the rotation operator which rotates vectors
around the $y'$ axis by $180^{\circ}$.
Obviously, $\mathcal{C}_{2y'}^2=1$ and $\mathcal{C}_{2y'}\mathbf{m}_{A(B)}^{\mathrm{eq}}=\mathbf{m}_{B(A)}^{\mathrm{eq}}$.
However, since $M_s^{A,0}\ne M_s^{B,0}$ and/or $d_A\ne d_B$, the joint operation of $\mathcal{C}_{2y'}$ and A-B sublayer exchange is no longer a symmetric operation of the entire SAF.
When a rf field with frequency $f=\omega/2\pi$ is excited, the magnetization vectors
deviate slightly from their equilibrium orientations and begin to vibrate.
We then expand $\mathbf{m}_{A(B)}=\mathbf{m}_{A(B)}^{\mathrm{eq}}+\delta\mathbf{m}_{A(B)} e^{\mathrm{i}\omega t}$
and $\mathbf{\tau}_{A(B)}=\mathbf{\tau}_{A(B)}^0 e^{\mathrm{i}\omega t}$.
In addition, we introduce $\delta\mathbf{m}_{\pm}\equiv\delta\mathbf{m}_A \pm \mathcal{C}_{2y'}\delta\mathbf{m}_B$ 
as the magnetization responses with even and odd parity (under $\mathcal{C}_{2y'}$) to the torque terms 
$\mathbf{\tau}_{\pm}=\mathbf{\tau}_A^0 \pm \mathcal{C}_{2y'}\mathbf{\tau}_B^0$.
By putting them into Eq. (\ref{Coupled_LLG_vectorial}) and keeping up to linear-order terms,
we get our central vectorial equation for the magnetization excitations $\delta\mathbf{m}_{\pm}$,
\begin{widetext} 
\begin{equation}\label{Coupled_LLG_EvenOddParity_general}
\begin{split}
\mathrm{i}\Omega\delta\mathbf{m}_{\pm}=&\mathbf{m}_{A}^{\mathrm{eq}}\times\left\{\frac{\lambda_{E}}{2}\left(M_s^A+M_s^B\right)\left(\delta\mathbf{m}_{\pm} \pm \mathcal{C}_{2y'}\delta\mathbf{m}_{\pm}\right)+\frac{\lambda_{E}}{2}\left(M_s^A-M_s^B\right)\left(\delta\mathbf{m}_{\mp} \pm \mathcal{C}_{2y'}\delta\mathbf{m}_{\mp}\right)\right.  \\ 
&\qquad\qquad
+\frac{M_s^{A,0}\pm M_s^{B,0}}{4}\left[\mathbf{e}_z\mathbf{e}_z\cdot+\mathcal{C}_{2y'}\mathbf{e}_z\left(\mathcal{C}_{2y'}\mathbf{e}_z\right)\cdot\right]\delta\mathbf{m}_+
+
\frac{M_s^{A,0}\pm M_s^{B,0}}{4}\left[\mathbf{e}_z\mathbf{e}_z\cdot-\mathcal{C}_{2y'}\mathbf{e}_z\left(\mathcal{C}_{2y'}\mathbf{e}_z\right)\cdot\right]\delta\mathbf{m}_-     \\
&\qquad\qquad
\left.+\frac{M_s^{A,0}\mp M_s^{B,0}}{4}\left[\mathbf{e}_z\mathbf{e}_z\cdot+\mathcal{C}_{2y'}\mathbf{e}_z\left(\mathcal{C}_{2y'}\mathbf{e}_z\right)\cdot\right]\delta\mathbf{m}_-
+
\frac{M_s^{A,0}\mp M_s^{B,0}}{4}\left[\mathbf{e}_z\mathbf{e}_z\cdot-\mathcal{C}_{2y'}\mathbf{e}_z\left(\mathcal{C}_{2y'}\mathbf{e}_z\right)\cdot\right]\delta\mathbf{m}_+\right\}+\mathbf{\tau}_{\pm},
\end{split}
\end{equation}
\end{widetext} 
where $\Omega=\omega/\gamma$.
Now $\delta\mathbf{m}_+$ and $\delta\mathbf{m}_-$ are coupled by both intrinsic (proportional to
$M_s^A-M_s^B$ and $M_s^{A,0}-M_s^{B,0}$) and extrinsic (proportional to $\mathcal{C}_{2y'}\mathbf{e}_z$) SB terms,
thus result in strong magnon-magnon interaction between the bare modes with even and odd parities.

Next we calculate the FMR spectroscopy.
After defining the average bare and ``thickness-modified" saturation magnetziation
as $\overline{M}_s^0=(M_s^{A,0}+M_s^{B,0})/2$ and $\overline{M}_s=(M_s^A+M_s^B)/2$, 
as well as their deviations 
$\kappa_0=(M_s^{A,0}-M_s^{B,0})/(M_s^{A,0}+M_s^{B,0})$
and $\kappa=(M_s^A-M_s^B)/(M_s^A+M_s^B)$ describing
the degree of asymmetry between the two FM sublayers, 
the secular equation for Eq. (\ref{Coupled_LLG_EvenOddParity_general}) is
\begin{widetext} 
\begin{equation}\label{secular_equation_general}
\left |\begin{array}{cccc}
p_1-\mathrm{i}\Omega & p_2 & p_3 & p_4   \\
q_1+2\lambda_{E}\overline{M}_s\cos^2\phi' & -p_1-\mathrm{i}\Omega & q_2+2\lambda_{E}\overline{M}_s\kappa\cos^2\phi' & -p_3   \\
p_3 & p_4-2\lambda_{E}\overline{M}_s\kappa & p_1-\mathrm{i}\Omega &  p_2-2\lambda_{E}\overline{M}_s  \\
q_2+2\lambda_{E}\overline{M}_s\kappa\sin^2\phi' & -p_3 & q_1+2\lambda_{E}\overline{M}_s\sin^2\phi' & -p_1-\mathrm{i}\Omega   \\
\end{array}\right|=0.
\end{equation}
\end{widetext} 
in which $p_1=\overline{M}_s^0(\cos\alpha\sin\phi'-\kappa_0\cos\beta\cos\phi')\cos\eta$,
$p_2=-\overline{M}_s^0\cos^2\eta$, 
$p_3=\overline{M}_s^0(\kappa_0\cos\alpha\sin\phi'-\cos\beta\cos\phi')\cos\eta$,
$p_4=-\overline{M}_s^0\kappa_0\cos^2\eta$, 
$q_1=[(p_1-p_3)^2(p_4+p_2)-(p_1+p_3)^2(p_4-p_2)]/[2(p_4^2-p_2^2)]$,
and $q_2=[-(p_1-p_3)^2(p_4+p_2)-(p_1+p_3)^2(p_4-p_2)]/[2(p_4^2-p_2^2)]$.
In addition, $\cos\alpha$, $\cos\beta$ and $\cos\eta$ are the three cosines 
of $\mathbf{e}_z$ in the new $x'y'z'$ coordinate system:
$\cos\alpha=(\cos\theta_A-\cos\theta_B)/(2\cos\phi')$,
$\cos\beta=(\cos\theta_A+\cos\theta_B)/(2\sin\phi')$,
and $\cos\eta=\sqrt{\sin^2\theta_A\sin^2\theta_B-(\cos\theta_A\cos\theta_B+\cos 2\phi')^2}/\sin 2\phi'$.
At last, $\phi'$ is the angle between $\mathbf{m}_A^{\mathrm{eq}}$ and $\mathbf{e}_{x'}$
which satisfies
$\cos 2\phi'=-\cos\theta_A\cos\theta_B+[(M_s^A\sin\theta_A)^2+(M_s^B\sin\theta_B)^2-(H\cos\psi/\lambda_{E})^2]/(2 M_s^A M_s^B)$.

In the most general case where intrinsic and extrinsic SB coexist ($\kappa_0\ne 0$, $\kappa\ne 0$,
$\psi>0$), Eq. (\ref{secular_equation_general}) can hardly be solved analytically.
However, based on the above results numerical calculation can always be performed once we 
have knowledge on $M_s^{A(B),0}$, $d_{A(B)}$, $\lambda_{E}$ and $\psi$.
By first fixing $H$ and solving the above secular equation then further sweeping $H$,
the entire FMR spectrum in the ``$\omega\sim H$" space can be obtained.
In Fig. 2 the dimensionless FMR spectra under different configurations are provided based on Eqs. (\ref{Coupled_LLG_EvenOddParity_general}) and (\ref{secular_equation_general}).
In all calculations, saturation magnetizations and magnetic fields are in the unit of $M_s^{A,0}$ 
and $\omega$ is in the unit of $\gamma M_s^{A,0}$.
In addition, $\lambda_{E}=0.1$.
In the absence of any SB, a mode crossing is clearly seen. 
When any single type of asymmetry appears, an anticrossing gap emerges.
While all kinds of SB coexist, the gap width is greatly enlarged.

\begin{figure} [htbp]
	\centering
	\includegraphics[width=0.48\textwidth]{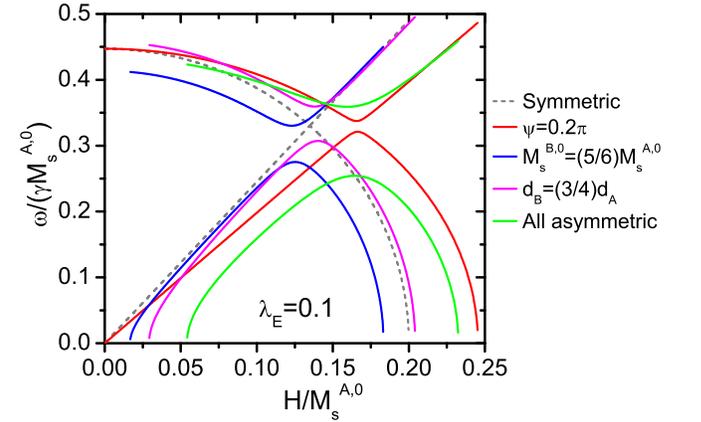}
	\caption{(Color online) Anticrossing gap opening when extrinsic and/or intrinsic symmetry breaking
		emerge. In all calculations based on Eqs. (\ref{Coupled_LLG_EvenOddParity_general}) and
		(\ref{secular_equation_general}), $\lambda_{E}=0.1$ and $M_s^{A,0}$ is taken as the unit of 
		saturation magnetizations and magnetic fields. $\omega$ is in the unit of 
		$\mu_0\gamma M_s^{A,0}$. The tilting angle ($\psi$) of dc fields and the ratios of bare 
		saturation magnetization ($M_s^{B,0}/M_s^{A,0}$) as well as sublayer thickness ($d_B/d_A$) are
		three sources of symmetry breaking. Each has been first independently shifted away from 
		the symmetrical configuration (red, blue and magenta curves), and then together (green curve).
	}\label{fig2}
\end{figure}

The mechanisms of inducing magnon-magnon coupling 
by extrinsic and intrinsic SB are different.
We have already described that for extrinsic SB, 
the breaking rotation symmetry of hard axis $\mathbf{n}$ (from magnetostatic interaction)
with respect to the pulled up rotation-symmetry axis (due to out-of-plane dc fields) is the basic reason.
While for intrinsic ones ($\psi=0$), the new rotation axis is still in the SAF plane
but the entire magnetic layout (including magnetization and magnetostatics) is no longer unchanged 
under the two-fold rotation.
The considerable entanglement between magnetization vibrations with opposite parity results in
the strong magnon-magnon coupling thus greatly affects the FMR spectroscopy of SAFs.

To acquire clearer physics meantime provide more convenient fitting tools for 
experiments, in the following we focus on a few special cases and present more 
details about the anticrossing gap.
In the first class, only extrinsic SB is present so that $M_s^{A,0}=M_s^{B,0}$, $d_A=d_B$ 
and $\psi>0$ [see Fig. 1(b)].
This is just the case similar to layered crystal $\mathrm{CrCl_3}$
and have been systematically investigated already, so we won't repeat it here. 
The second class includes the cases where only intrinsic SB is involved,
which means only in-plane dc fields are applied ($\psi=0$) as depicted in Fig. 1(c).
As a result, $\mathbf{e}_{z'}\equiv \mathbf{e}_z$ thus the new $x'y'$ plane is identical to the old $xy$ one.
Now the AFM and FM critical fields take much simpler forms: $H_{\mathrm{AFM}}=\lambda_{E}|M_s^A-M_s^B|$
and $H_{\mathrm{FM}}=\lambda_{E}(M_s^A+M_s^B)$.
Consequently, the secular equation (\ref{secular_equation_general}) is simplified to
\begin{equation}\label{secular_equation_general_intrinsic}
\tilde{\omega}^4-\left(\tilde{h}^2+\mu\frac{1+\kappa_0\kappa}{2\lambda_{E}}\right)\tilde{\omega}^2+\frac{2\lambda_{E}+\nu}{4\lambda_{E}^2}\left(1-\tilde{h}^2\right)\left(\tilde{h}^2-\kappa^2\right)=0,
\end{equation}
with $\tilde{\omega}\equiv\Omega/H_{\mathrm{FM}}$, $\kappa<\tilde{h}\equiv H/H_{\mathrm{FM}}<1$,
$\mu=(M_s^{A,0}+M_s^{B,0})/(M_s^A+M_s^B)$ and $\nu=(M_s^{A,0}/M_s^A)\cdot(M_s^{B,0}/M_s^B)$.
Generally, a gap appears as long as $\kappa_0^2+\kappa^2\ne 0$.

In particular, this gap can be further analyzed for two special cases.
In the first case, the two FM sublayers have the same thickness ($d_A=d_B$) but
are made of different FM materials ($M_s^{A,0}\ne M_s^{B,0}$).
Then $M_s^{A(B)}= M_s^{A(B),0}$, thus $\mu=\nu=1$ and $\kappa_0=\kappa\ne 0$.
Similar to MacNeill \textit{et. al.} in 2019\cite{MacNeill_PRL_2019}, 
Now the secular equation (\ref{secular_equation_general_intrinsic})
can be rewritten into the eigenvalue problem of a $2\times2$ matrix as
\begin{equation}\label{Matrix_2_by_2_omega_a_o}
\left |\begin{array}{cc}
\tilde{\omega}_{\mathrm{a}}^2(\tilde{h})-\tilde{\omega}^2 & \tilde{\Delta}^2   \\
\tilde{\Delta}^2 & \tilde{\omega}_{\mathrm{o}}^2(\tilde{h})-\tilde{\omega}^2   \\
\end{array}\right|=0.
\end{equation}
Here $\tilde{\omega}_{\mathrm{a}}=\sqrt{1+(2\lambda_{E})^{-1}}\tilde{h}$ and
$\tilde{\omega}_{\mathrm{o}}=[(1+\kappa^2)(1-\tilde{h}^2)/(2\lambda_{E})+(\kappa^2\tilde{h}^2)/(2\lambda_{E})]^{1/2}$ are the bare in-phase (acoustic) and out-of-phase (optical) mode frequencies, respectively.
Meantime $\tilde{\Delta}=[(2\lambda_{E}+1)\kappa^2/(4\lambda_{E}^2)]^{1/4}$ describes the 
\emph{dc-field-independent} magnon-magnon coupling strength.
When $|\kappa|\ll 1$ (nearlly symmetric), 
$\tilde{\Delta}$ is negligible thus the solution of Eq. (\ref{Matrix_2_by_2_omega_a_o}) are approximately $\tilde{\omega}\approx \tilde{\omega}_{\mathrm{a}}$ and
$\tilde{\omega}\approx \tilde{\omega}_{\mathrm{o}}$.
For finite $\kappa$, when $\tilde{h}$ is close to $\kappa$ or 1,
the solutions of Eq. (\ref{Matrix_2_by_2_omega_a_o}) only deviate slightly from 
$\tilde{\omega}_{\mathrm{a}}$ and $\tilde{\omega}_{\mathrm{o}}$.
When the optical and acoustic modes get closer in frequency, they will be hybridized by
$\tilde{\Delta}$ term and thus a gap is opened.
Direct calculations yield that this gap is an indirect gap (see Supplement Material):
the minimum $\tilde{\omega}_{\mathrm{up}}^{\mathrm{min}}$ 
(maximum $\tilde{\omega}_{\mathrm{down}}^{\mathrm{max}}$) of the ``up (down)"
branch takes place at $\tilde{h}_{\mathrm{up}}^{\mathrm{min}}$ ($\tilde{h}_{\mathrm{down}}^{\mathrm{max}}$),
where $\tilde{h}_{\mathrm{up}}^{\mathrm{min}}\ne \tilde{h}_{\mathrm{down}}^{\mathrm{max}}$ for nozero $\kappa$.
The corresponding dimensionless gap width reads $\delta\tilde{\omega}=|\kappa|\sqrt{(2\lambda_{E}+1)/[\lambda_{E}(\lambda_{E}+1)]}$.
Interestingly, the real gap width $\delta\tilde{\omega}$ is linear
to $|\kappa|$ (degree of asymmetry between the two FM sublayers),
which is different from the square-root dependence of the coupling $\tilde{\Delta}$ on $\kappa$.
By recalling the definitions of $\tilde{\omega}$ and $\kappa$,
we then get the dimensional gap width
\begin{equation}\label{Same_thickness_gap_width_dimensional}
\delta f=\sqrt{\frac{\lambda_{E}\left(2\lambda_{E}+1\right)}{\lambda_{E}+1}}\frac{\gamma}{2\pi}|M_s^{A,0}-M_s^{B,0}|,
\end{equation}
which should be more useful for experimental physicists. 
Obviously, the AFM interlayer coupling and the inhomogeneity of sublayers are both crucial for the gap opening.

Another interesting case is that the two FM sublayers are made of the same material ($M_s^{A,0}=M_s^{B,0}$)
but have different thickness ($d_A\ne d_B$), which is the most common choice in real experiments.
Accordingly $\kappa_0=0$, $\kappa=(d_A-d_B)/(d_A+d_B)$ and $\mu=\nu=1-\kappa^2$.
Now Eq. (\ref{secular_equation_general_intrinsic}) can not be reorganized to the $2\times2$ matrix form as that
in Eq. (\ref{Matrix_2_by_2_omega_a_o}) if we require the coupling term to be $\tilde{h}$-independent.
However, similar calculus shows that now the gap is also indirect.
The new extremum, extremum locations and the gap width become more complicated (see Supplement Material),
and we then focus on the situation where $|\kappa|\ll 1$.
After standard linearization operation, the gap width is approximated to
another linear dependence on $|\kappa|$ as
$|\kappa|[(2\lambda_{E}+1)/(\lambda_{E}+1)]^{3/2}/(2\sqrt{\lambda_{E}})$.
Back to dimensional form, we have
\begin{equation}\label{Same_material_gap_width_dimensional}
\delta f'\approx\frac{\sqrt{\lambda_{E}}}{2}\left(\frac{2\lambda_{E}+1}{\lambda_{E}+1}\right)^{\frac{3}{2}}\frac{|1-(d_B/d_A)^2|}{2 (d_B/d_A)}\frac{\gamma}{2\pi} M_s^{A,0}.
\end{equation}

Eq. (\ref{Same_material_gap_width_dimensional}) can be directly compared with existing 
experimental measurements for asymmetrical NiFe/Ru/NiFe SAFs\cite{Belmeguenai_JPCM_2008,Liu_PRB_2014}.
For NiFe(13.6 nm)/Ru($t_{\mathrm{Ru}}$)/NiFe(27.2 nm) SAFs in Ref. \cite{Belmeguenai_JPCM_2008}, 
we choose $t_{\mathrm{Ru}}=4.7\ \mathrm{\mathring{A}}$ as an example.
By taking $M_s^{A,0}=860\mathrm{\ kA\ m^{-1}}$ and 
$J_{\mathrm{IEC}}\approx |J_1|=286\ \mathrm{\mu J\ m^{-2}}$,
we get $\lambda_{E}\approx 0.015$ and $\delta f'\approx 1.4 \mathrm{\ GHz}$.
While for $\mathrm{Ni_{80}Fe_{20}(200\ \mathring{A})}$/Ru($t_{\mathrm{Ru}}$)/$\mathrm{Ni_{80}Fe_{20}(100\ \mathring{A})}$ SAFs in Ref. \cite{Liu_PRB_2014}, 
the magnetic parameters for $t_{\mathrm{Ru}}=3.3\ \mathrm{\mathring{A}}$ are 
$M_s^{A,0}=720\mathrm{\ kA\ m^{-1}}$ and $J_{\mathrm{IEC}}\approx 154\ \mathrm{\mu J\ m^{-2}}$.
These lead to $\lambda_{E}\approx0.016$ and $\delta f'\approx 1.2 \mathrm{\ GHz}$.
For other Ru thickness, similar calculations can be performed and all results show
good agreement between analytics and experimental data.

Interestingly, in FeCoB/Ru/FeCoB SAFs we can acquire larger $\lambda_{E}$, although
nearly all existing published experiments are performed in symmetrical cases\cite{Shiota_PRL_2020,Shiota_SciAdv_2020,Sud_PRB_2020,Waring_PRApplied_2020}.
For symmetrical FeCoB thickness being $15\ \mathrm{nm}$\cite{Shiota_PRL_2020,Shiota_SciAdv_2020}, 
$3\ \mathrm{nm}$\cite{Sud_PRB_2020} and $5\ \mathrm{nm}$\cite{Waring_PRApplied_2020},
the respective $\lambda_{E}$ are estimated to be $0.033$, $0.093$ and $0.119(0.141)$ (two samples therein).
Combined with larger saturation magnetization of FeCoB, asymmetrical FeCoB/Ru/FeCoB SAFs
are expected to open larger indirect gaps.
In view of this, we have performed related measurements in asymmetrical CoFeB/Ir/CoFeB SAFs 
(experimental details will be published elsewhere).
In the ``CoFeB(10 nm)/Ir(0.6 nm)/CoFeB(12 nm)" SAF,
direction measurements provide $M_s^{A,0}=1400\mathrm{\ kA\ m^{-1}}$ and 
$H_{\mathrm{ex}}\equiv 2J_{\mathrm{IEC}}/[\mu_0 M_s^{A,0}(d_A+d_B)]=549\ \mathrm{Oe}$,
thus $\lambda_{E}\approx 0.0312$ and $\delta f'\approx 0.80 \mathrm{\ GHz}$, which is
in good agreement with experimental observation of gap width (0.74 GHz). 
In addition, the gap is observed to take place around 760 Oe, which is
also in accordance with theoretical prediction (770 Oe).
We are looking forward to more experimental measurements in the near future.
In very recent symmetrical CoFeB/Ru/CoFeB SAFs\cite{Sud_PRB_2020}, out-of-plane
dc fields are needed to open gaps up to 1 GHz. 
Our results provide the possibility that by appropriately designing the thickness ratio of two CoFeB sublayers,
greater gap can be opened which indicates stronger magnon-magnon coupling.
On the other hand, we know that it is very difficult to prepare inhomogeneous SAFs ($M_s^{A,0}\ne M_s^{B,0}$)
experimentally,
however this issue is indeed worth exploring further and Eq. (\ref{Same_thickness_gap_width_dimensional})
should help to reveal interesting physics.

At the end of this Letter, several points need to be clarified.
First, in this work the crystalline anisotropy has been neglected because of two reasons:
(i) in most existing experimental setups, the FM sublayers of SAFs are 
made from soft magnetic materials which can be viewed as isotropic;
(ii) the explicit orientations of $\mathbf{m}_A^{\mathrm{eq}}$ and $\mathbf{m}_B^{\mathrm{eq}}$
can hardly be obtained analytically if the in-plane crystalline anisotropy is considered 
(even for the simplest uniaxial case),
but they are crucial for obtaining the 
vectorial LLG equation and then the secular equations for FMR frequencies.
For the above reasons, in this work we choose the isotropic case for simplicity,
but it can cover the vast majority of experimental data.
Note that our analytics also holds for perpendicular-magnetic-anisotropy case
as long as we change $M_s^{A(B),0}$ to $M_s^{A(B),0}-H_K^{A(B)}$ in Eq. (\ref{Coupled_LLG_vectorial})
where $H_K^{A(B)}$ is the out-of-plane anisotropic field in the respective sublayer.
Another neglected term is the Gilbert damping term.
In most investigations of spin wave, the damping term is dropped off when 
only the resonance spectrum is under consideration.
However when the linewidth is also of interest, then the damping process
should be included in Eq. (\ref{Coupled_LLG_EvenOddParity_general}) by a respective term 
$\mathrm{i}\omega\alpha\mathbf{m}_A^{\mathrm{eq}}\times\delta\mathbf{m}_{\pm}$
with $\alpha$ being the Gilbert damping coefficient\cite{MacNeill_PRL_2019}.

In summary, we have proposed a simple but revealing theory for understanding
the rich structure of FMR spectroscopy in asymmetrical SAFs under tilting dc magnetic fields.
Both intrinsic and extrinsic SB cause entanglement between magnetization vibrations
with opposite parity, thus excite strong magnon-magnon coupling between 
the bare in-phase and out-of-phase modes, then eventually result in 
the anticrossing gap in microwave absorption spectroscopy.
In addition, for pure intrinsic SB the two-fold rotation symmetry of both
the magnetization and magnetostatics around $\mathbf{m}_A^{\mathrm{eq}}+\mathbf{m}_B^{\mathrm{eq}}$
fails. While for pure extrinsic SB, only that of magnetostatics fails.
This new picture helps to understand the rich experimental data of FMR spectra for 
existing SAFs and future measurements on other SAF or SAF-like systems.

J.L. acknowledges support from the Natural Science Foundation for Distinguished Young Scholars of
Hebei Province of China (A2019205310).
M.L. is funded by the National Natural Science Foundation of China (Grant No. 11947023),
the Project of Hebei Province Higher Educational Science and Technology Program (QN2019309)
and the PhD Research Startup Foundation of Shijiazhuang University (20BS022).
W.H. is supported by the National Natural Science Foundation of China (Grant No. 51871235).


\end{document}